\documentclass[aps,prl,epsfig,showpacs,superscriptaddress,twocolumn]{revtex4}

\usepackage{graphicx,amsmath,amsfonts,textcomp}

\begin{document}

\title{Mechanisms of exchange bias with multiferroic BiFeO$_3$ epitaxial thin films}

\author{H. B\'ea}
\affiliation{Unit\'e Mixte de Physique CNRS-Thales, Route d\'epartementale 128, 91767 Palaiseau, France}

\author{M. Bibes}
\affiliation{Institut d'Electronique Fondamentale, CNRS, Universit\'e Paris-Sud, 91405 Orsay, France}

\author{F. Ott}
\affiliation{Laboratoire L\'eon Brillouin CEA/CNRS UMR12, Centre d'Etudes de Saclay, 91191 Gif sur Yvette,
France}

\author{B. Dup\'e}
\affiliation{Unit\'e Mixte de Physique CNRS-Thales, Route d\'epartementale 128, 91767 Palaiseau, France}

\author{X.-H. Zhu}
\affiliation{Unit\'e Mixte de Physique CNRS-Thales, Route d\'epartementale 128, 91767 Palaiseau, France}

\author{S. Petit}
\affiliation{Laboratoire L\'eon Brillouin CEA/CNRS UMR12, Centre d'Etudes de Saclay, 91191 Gif sur Yvette,
France}

\author{S. Fusil}
\affiliation{Unit\'e Mixte de Physique CNRS-Thales, Route d\'epartementale 128, 91767 Palaiseau, France}

\author{C. Deranlot}
\affiliation{Unit\'e Mixte de Physique CNRS-Thales, Route d\'epartementale 128, 91767 Palaiseau, France}

\author{K. Bouzehouane}
\affiliation{Unit\'e Mixte de Physique CNRS-Thales, Route d\'epartementale 128, 91767 Palaiseau, France}

\author{A. Barth\'el\'emy}
\email{agnes.barthelemy@thalesgroup.com} \affiliation{Unit\'e Mixte de Physique CNRS-Thales, Route d\'epartementale 128, 91767 Palaiseau, France}

\date{\today}

\begin{abstract}

\vspace{0.3cm}

We have combined neutron scattering and piezoresponse force microscopy to study the relation between the
exchange bias observed in CoFeB/BiFeO$_3$ heterostructures and the multiferroic domain structure of the
BiFeO$_3$ films. We show that the exchange field scales with the inverse of the ferroelectric and
antiferromagnetic domain size, as expected from Malozemoff's model of exchange bias extended to multiferroics.
Accordingly, polarized neutron reflectometry reveals the presence of uncompensated spins in the BiFeO$_3$ film
at the interface with the CoFeB. In view of these results we discuss possible strategies to switch the
magnetization of a ferromagnet by an electric field using BiFeO$_3$.

\end{abstract}

\maketitle

The renaissance of multiferroics \cite{spaldin2005,eerenstein2006}, i.e. materials in which at least two ferroic
or antiferroic orders coexist, is motivated by fundamental aspects as well as their possible application in
spintronics \cite{bibes2007}. Such compounds are rare and the very few that possess simultaneously a finite
magnetization and polarization usually order below about 100K \cite{kato82,yamasaki2006,gajek2007}.
Ferroelectric antiferromagnets (FEAF) are less scarce, and some exhibit a coupling between their two order
parameters. This magnetoelectric (ME) coupling allows the reversal of the ferroelectric (FE) polarization by a
magnetic field \cite{kimura2003c} or the control of the magnetic order parameter by an electric field
\cite{zhao2006}.

The practical interest of conventional antiferromagnets (AF) is mainly for exchange bias in spin-valve
structures. The phenomenon of exchange bias (EB) \cite{meiklejohn56} manifests itself by a shift in the
hysteresis loop of a ferromagnet (FM) in contact with an AF and arises from the exchange coupling at the FM/AF
interface \cite{nogues99,radu2007}. Combining this effect with the ME coupling in a FEAF/FM bilayer can allow
the reversal of the FM magnetization via the application of an electric field through the FEAF, as reported
recently at 2K in YMnO$_3$/NiFe structures \cite{laukhin2006}.

To exploit these functionalities in devices one needs to resort to FEAF materials with high transition
temperatures. BiFeO$_3$ (BFO) is a FE perovskite with a Curie temperature of 1043K \cite{teague70} that orders
antiferromagnetically below T$_N$=643K (T$_N$: N\'eel temperature) \cite{kiselev62}. BFO thin films have a very
low magnetization ($\sim$0.01 $\mu_B$/Fe) compatible with an AF order \cite{bea2005,bea2006b}, and remarkable FE
properties with polarization values up to 100 $\mu$C.cm$^{-2}$ range \cite{lebeugle2007}. Recently, we reported
that BFO films can be used to induce an EB on adjacent CoFeB layers at room temperature \cite{bea2006d}. This
observation together with the demonstration of a coupling between the AF and FE domains \cite{zhao2006} paves
the way towards the room-temperature electrical control of magnetization with BFO. However, several questions
remain before this can be achieved. Key issues concern the precise magnetic structure of BFO thin films, and the
mechanisms of EB in BFO-based heterostructures.

In this Letter, we report on the determination of the magnetic structure of BFO films by means of neutron
diffraction (ND), and the analysis of the EB effect in CoFeB/BFO heterostructures in terms of Malozemoff's model
\cite{malozemoff87}. Accordingly, we find a clear dependence of the amplitude of the exchange field H$_E$ with
the size of the multiferroic domains, which provides a handle to control the magnetization of the CoFeB film by
an electric field. The observation of EB and enhanced coercivity correlates with the presence of uncompensated
spins at the interface between the FM and the AF, as detected by polarized neutron reflectometry (PNR).

BiFeO$_3$ films were grown by pulsed laser deposition \cite{bea2005}, directly onto (001)- or (111)-oriented
SrTiO$_3$ (STO) or (001)-oriented LaAlO$_3$ (LAO) substrates, or onto 10-25 nm-thick metallic buffers of
La$_{2/3}$Sr$_{1/3}$MnO$_3$ (LSMO) or SrRuO$_3$ (SRO) \cite{bea2006c}. 7.5 nm-thick CoFeB layers were sputtered
in a separate chamber at 300K in a magnetic field of 200 Oe, after a short plasma cleaning. The samples were
capped by 10-30 nm of Au. High resolution x-ray diffraction evidenced a cube-on-cube growth for all the
perovskite layers onto the substrates. CoFeB was amorphous. While the (111)-oriented films were found to be
rhombohedral as bulk BFO \cite{bea2007c}, (001)-oriented films were found tetragonal or monoclinic
\cite{bea2007}.

A first key information that is usually required to analyze EB is the magnetic structure of the AF. Bulk BFO is
known to have a G-type AF order \cite{kiselev62}, with a superimposed cycloidal modulation \cite{sosnowska82}.
In view of the strong strain sensitivity of the properties of FE and magnetic oxides, one can anticipate that
the magnetic order of BFO films might be different from that of the bulk. In order to determine their magnetic
structure, selected (001)- and (111)-oriented BFO films were thus characterized by ND with the triple axis 4F1
spectrometer at the Orph\'ee reactor of the Laboratoire L\'eon Brillouin (LLB) \cite{bea2007c}.

\begin{figure}[h!]
\centering
\includegraphics[width=\columnwidth]{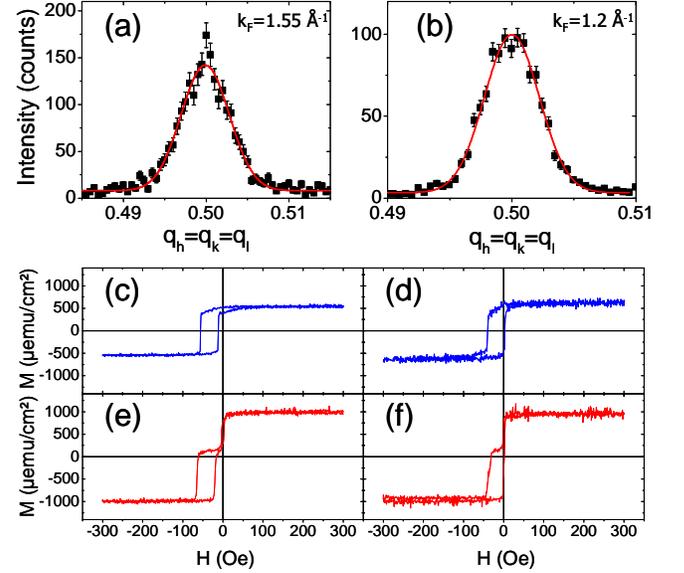}
\caption{(a) and (b) Neutron diffraction scans close to the [\textonehalf \, \textonehalf \, \textonehalf]
reflection for (001)- and (111)-oriented BFO films on STO, at 300K. Magnetic field dependence of the
magnetization of BFO/CoFeB (c,d) and LSMO/BFO/CoFeB samples (e,f) grown on (001)-oriented STO (c,e) and
(111)-oriented STO (d,f), at 300K.} \label{neutrons-mh}
\end{figure}

In a G-type AF, superstructure peaks are expected to appear at [\textonehalf \ \textonehalf \ \textonehalf]-type
reflections. In Fig. \ref{neutrons-mh}a and b, we show the diffracted intensity at the [\textonehalf \,
\textonehalf \, \textonehalf] reflection in BFO films grown on (001)- and (111)-oriented STO. Clearly an AF peak
is present for both films. On the other hand, no intensity was measured at [0 0 \textonehalf]-type or
[\textonehalf \, \textonehalf \, 0]-type reflections, characteristic of A-type and C-type antiferromagnetism,
respectively. These data thus show that both (001)- and (111)-oriented films are antiferromagnetic with a G-type
order similar to that of the bulk. In other words, neither strain nor changes in the unit-cell symmetry modify
the type of magnetic order, besides destroying the cycloidal modulation (see Ref. \onlinecite{bea2007} for
details on this aspect).

Within a simplistic model of EB, the exchange field H$_E$ depends on the interface coupling J$_{eb}$=$J_{ex} S_F
S_{AF}/a^2$ (J$_{ex}$ is the exchange parameter, S${_F}$ and S$_{AF}$ are the spin of the interfacial atoms in
the ferromagnet and the AF, respectively, and $a$ is the unit cell parameter of the AF), on the magnetization
and thickness of the ferromagnet M$_{FM}$ and t$_{FM}$, and on the anisotropy and the thickness of the AF
K$_{AF}$ and t$_{AF}$ \cite{radu2007}.

\vspace{-1.4em}
\begin{equation}\label{he}
\begin{split}
H_E & = -\frac{J_{eb}}{\mu_0 M_{FM} t_{FM}} \sqrt{1-\frac{J_{eb}^2}{4 K_{AF}^2 t_{AF}^2}} \\
& = H_{E}^\infty\sqrt{1-\frac{1}{4\Re^2}}
\end{split}
\end{equation}

\noindent provided that 1/${4\Re^2}$ is smaller than 1 (otherwise H$_E$ is zero). If t$_{AF}$ is large, $\Re
>> 1$ and thus H$_E \simeq H_{E}^\infty$. In a naive picture of perfect surfaces and because the
AF order is G-type in both (001) and (111) films, the (111) films are expected to have magnetically
uncompensated surfaces, yielding S$_{AF}$=5/2, while the (001) films should have compensated surfaces, yielding
in average S$_{AF}$=0. Therefore, a finite H$_E$ should be found for (111) films only. For (111) films, taking
J$_{ex}$=5 10$^{-22}$ J \cite{ruette2004}, a=3.96 $\rm{\AA}$, S$_F$=1/2, S$_{AF}$=5/2, M$_{FM}$=800 kA/m and
t$_{FM}$=7.5 nm one can estimate J$_{eb}$=4 mJ/m$^2$ and the large value of H$_E$=6.5 kOe.

Figure \ref{neutrons-mh}c and d show M(H) hysteresis loops measured at 300K for BFO(70nm)/CoFeB stacks grown on
STO(001) and STO(111). The loops are shifted towards negative magnetic field values by an exchange field H$_E$
of -39 Oe for (001) films and -19 Oe for (111) films. Furthermore the loops are enlarged by some tens of Oe
compared to those measured on CoFeB single films \cite{bea2006d}. From M(H) data for LSMO/BFO/CoFeB samples (see
Fig. \ref{neutrons-mh}e and f), a symmetric (i.e. not showing EB) contribution from the LSMO films is visible
(the Curie temperature of our LSMO films is about 330K) in addition to that coming from the CoFeB. For a given
thickness, the exchange field experienced by the CoFeB is virtually the same irrespective of the presence of the
LSMO buffer layer.

Malozemoff's random field model \cite{malozemoff87} has been proposed to resolve the long-standing discrepancy
between the model of Eq.(\ref{he}) and the experimental data \cite{nogues99}. It considers that in the presence
of some atomic-scale disorder that locally creates a net magnetization in the AF at the interface with the FM,
the AF splits into domains, which decreases considerably the interface coupling that now writes:

\vspace{-1.3em}
\begin{equation}\label{jeb-malo}
J_{eb}=\frac{\zeta J_{ex} S_{AF} S_F}{a L}
\end{equation}

L is the antiferromagnetic domain size and $\zeta$ a factor depending on the shape of the domains and on the
average number $z$ of frustrated interaction paths for each uncompensated surface spin \cite{malozemoff87}. For
hemispherical bubble domains, $\zeta \simeq 2z/ \pi$ and $z$ is of order unity \cite{malozemoff87}. From
equations (\ref{he}) and (\ref{jeb-malo}) and for large AF thickness ($\Re
>> 1$) the exchange field should vary as $1/L$:

\vspace{-1.3em}
\begin{equation}\label{he-malo}
H_E = H_E^\infty = -\frac{\zeta J_{ex} S_{AF} S_F}{\mu_0 M_{FM} t_{FM} a L}
\end{equation}

\noindent and indeed, a linear dependence of H$_E$ with 1/L was experimentally observed in several AF/FM systems
\cite{takano97,scholl2004}, providing strong support to the model.

In order to determine the domain size in antiferromagnets, a technique of choice is X-ray photoelectron emission
microscopy \cite{stohr99,scholl2000}. Alternatively, an estimation of the average domain size can be inferred
from the width of the ND peaks that reflects the coherence length in the sample \cite{borchers2000}. In some
multiferroics like BFO, the FE and AF domains are coupled \cite{zhao2006} so that it is possible to infer the
size and distribution of the AF domains by imaging the FE domains, e.g. using piezoresponse force microscopy
(PFM).

We have characterized the FE domains in two sets of BFO samples by combining in-plane and out-of-plane PFM
measurements \cite{catalan2007}. A first set consists of $\sim$65 nm-thick BFO films grown on different buffers
and substrates (see table \ref{tableau}). A second set consists of BFO films with varying thickness (5 nm $\leq
t_{AF} \leq$ 240 nm) grown on LSMO//STO(001). In BFO, since the polarization is oriented along the $<$111$>$
directions there generally exist 8 possible polarization orientation variants. This leads to a large number of
possible domain patterns and to several types of domain walls (DWs) depending on the angle between the
polarization vectors in the adjacent domains (71, 109 or 180$^\circ$). In principle only 71$^\circ$ and
109$^\circ$ FE DWs correspond to an AF DW \cite{zhao2006}. In most of the samples considered in this study, the
three types of DWs are present, with the density of 180$^\circ$ type DWs being negligible.

In the following we analyze the exchange field in terms of the average FE domain size L$_{FE}$ that we identify
to half the FE domain periodicity, and the average AF domain size L$_{AF}$ that we identify to the coherence
length in the ND experiments. Even in the case of a strict correspondence between the FE and AF domains,
L$_{FE}$ is expected to be larger than L$_{AF}$ because it comprises both the domain and the DW widths while
L$_{AF}$ mostly reflects the domain width.

\vspace{-1em}
\begin{table} [!h]
 \caption{Experimental values of the exchange field and ferroelectric domain size for samples of the first set.} \label{tableau}
 \vspace{0.5em}
\begin{ruledtabular}
\begin{tabular}{ccccc}

Substrate  & ~ Buffer ~ & $t_{AF}$(nm) & ~ H$_E$ ~ &  L$_{FE}$(nm) ~ \\

STO(001) & ~ LSMO ~ & 70 & ~ -39  ~ & ~ 58 ~\\
STO(001) & ~ SRO ~ & 70 & ~ -14.5  ~ & ~ 98 ~\\
STO(111) & ~ LSMO ~ & 70 & ~ -19  ~ & ~ 68 ~\\
STO(111) & ~ SRO ~ & 70 & ~ -39  ~ & ~ 48 ~\\
LAO(001) & ~ LSMO ~ & 60 & ~ 0  ~ & ~ 350 ~\\
LAO(001) & ~ SRO ~ & 60 & ~ -29  ~ & ~ 55 ~\\

\vspace{-1em}

\end{tabular}
\end{ruledtabular}
\end{table}

\begin{figure}
\centering
\includegraphics[width=\columnwidth]{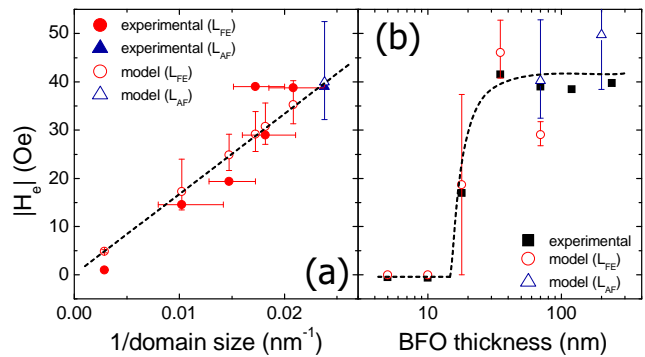}
\caption{(a) Dependence of the exchange field on the inverse of the domain size for several $\sim$65 nm thick
BFO films. (b) Thickness dependence of the exchange field for CoFeB/BFO stacks grown on STO(001). The dotted
lines are guides to the eye.} \label{dep}
\end{figure}

Let us first compare the values of L$_{AF}$ and L$_{FE}$ for the samples of Fig. \ref{neutrons-mh}. From Fig.
\ref{neutrons-mh}a we calculate L$_{AF}$=42 nm while from PFM we find L$_{FE}$=58 nm for this sample. For the
sample of Fig. \ref{neutrons-mh}b, L$_{AF}$=56 nm and L$_{FE}$=98 nm. L$_{FE}$ is larger than L$_{AF}$ by some
tens of nm that likely correspond to the DW width \cite{DWwidth}. When L$_{AF}$ increases L$_{FE}$ increases
also, as expected if the AF and FE domains are correlated, as found by Zhao \emph{et al} \cite{zhao2006}.

In figure \ref{dep}a we plot the exchange field as a function of the inverse of the domain size for the first
set of samples (constant thickness). A linear variation of H$_E$ with $1/L$ is observed, as expected from Eq.
(\ref{he-malo}). Furthermore, there is an excellent \emph{quantitative} agreement between the model and the data
as illustrated by the similarity between the experimental points (solid symbols in Fig. \ref{dep}a) and the
values of H$_E$ calculated using the domain sizes (open symbols; the only free parameter is $\zeta$ that we set
to 3.2).

As illustrated by Fig. \ref{dep}b, this model also accounts for the thickness dependence of the exchange field.
Below a critical BFO thickness of about 10 nm, there is no exchange bias but at larger thickness, H$_E$
increases abruptly and takes values of about 40 Oe for t$_{AF} \geq 35$ nm. A similar thickness dependence was
reported for other AF/FM systems such as FeMn/NiFe \cite{jungblut94} and is expected from Eq. (\ref{he}).
Combining Eq. (\ref{he}) and Eq. (\ref{jeb-malo}), and using the measured domain sizes one can calculate H$_E$
for these films. The only free parameters are $\zeta$ and K$_{AF}$. As shown by the open symbols in Fig.
\ref{dep}b, a rather good agreement with the data is obtained for $\zeta$=3.2 (as before) and K$_{AF}$=6.8
kJ/m$^3$. We note that this value of K$_{AF}$ is lower than the value inferred for the bulk from electron spin
resonance experiments \cite{ruette2004} by about one order of magnitude. This can be due to the low thickness of
the BFO films and possibly to strain effects, or reflect modified magnetic properties at the BFO/CoFeB
interface, as will be discussed in the following.

\begin{figure}
\centering
\includegraphics[width=\columnwidth]{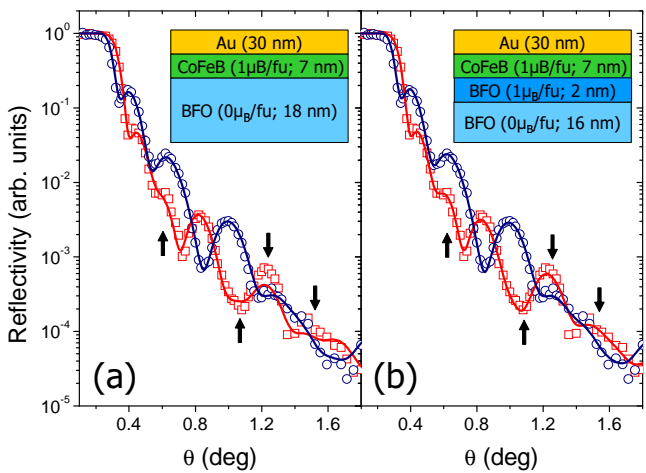}
\caption{Polarized neutron reflectivity data measured in a saturating field of 1.2 T (R$_{++}$: squares;
R$_{--}$: circles) fitted with a three-layer model (a) and a four-layer model (b). Introducing an interface
layer of $\sim$2 nm carrying a moment of 1$\pm{0.5}$ $\mu_B$/f.u. significantly improves the quality of the fit
for up-up reflectivity (see arrows).} \label{Refl}
\end{figure}

To validate the analysis of our data with Malozemoff's model, we have attempted to detect the presence of a net
magnetization in BFO close to the interface with CoFeB using PNR. The PNR measurements were carried out with the
PRISM instrument of the LLB at room temperature. Spin-up and spin-down reflectivities (R$_{++}$ and R$_{--}$)
were collected and the data were corrected for the polarization efficiency. The least square fittings were made
using the Simul-Reflec software. The structural parameters of the layers were first determined by fitting X-ray
reflectometry data (not shown). In the following we discuss the results for a
Au(30nm)/CoFeB(7.5nm)/BFO(18nm)//STO(001) sample displaying an EB of 18 Oe and in which the CoFeB/BFO interface
has a roughness of only 0.5 nm, yielding high quality reflectivity data.

Figure \ref{Refl} shows the PNR results and the corresponding fits using two different sample models. In the
first sample model (Fig. \ref{Refl}a) we only consider the presence of three layers, i.e. Au, CoFeB and BFO. The
best fit is obtained for the structure Au(31.0$\pm{0.5}$nm)/CoFeB(7.0$\pm{0.5}$nm)/BFO(18$\pm{0.5}$nm) with zero
magnetization in the BFO and Au layers, and a magnetization of 1$\pm{0.05}$ $\mu_B$/f.u. for the CoFeB. As can
be appreciated, the fit is good but not perfect (see arrows). The fit quality is significantly improved,
especially at high $\theta$, by using a four-layer model, i.e. splitting the BFO layer in two. The best fit is
then obtained if a 2$\pm{0.5}$nm layer carrying a magnetic moment of 1$\pm{0.5}$$\mu_B$/f.u. is present in the
BFO, at the interface with the CoFeB. This ultrathin magnetic layer inside the BFO accounts for the presence of
a large density of uncompensated spins (corresponding to a surface moment m$_s$=31.8 $\mu_B$.nm$^{-2}$) at the
BFO/CoFeB interface, similarly to results reported on the Co/LaFeO$_3$ exchange bias system \cite{hoffmann2002}.

The expected surface moment due to pinned uncompensated spins within Malozemoff's model is
m$_s^{pin}$=2S$_{AF}$/aL$\simeq$0.32 $\mu_B$.nm$^{-2}$, which represents only a small fraction of the surface
moment measured by PNR. The majority of uncompensated spins is thus unpinned (as observed by X-ray magnetic
circular dichroism in Co/IrMn \cite{ohldag2003}) and rotates with the CoFeB, producing an increase of the
coercive field, as found experimentally (see Fig. \ref{neutrons-mh}). This observation suggests that two
different, yet possibly related and complementary, strategies are possible to tune the magnetic switching fields
of the ferromagnet electrically. One would rely on the manipulation of the pinned uncompensated spins to modify
the exchange field H$_E$, e.g. by changing the domain size that should be controllable by ad-hoc electrical
writing procedures. The other could consist in controlling the unpinned uncompensated spins in order to alter
the coercive field H$_C$. This might be achieved by modifying the effective surface anisotropy of the AF, for
instance by playing with the ferroelastic energy of the domains, which would change the magnetoelastic
contribution to the anisotropy.

In summary, we analyzed the exchange bias in the CoFeB/BiFeO$_3$ system and found that the exchange field does
not correlate with the type of magnetic surface of the antiferromagnet - compensated or uncompensated. Rather,
the exchange field scales with the inverse of the ferroelectric and antiferromagnetic domain size in the
multiferroic BiFeO$_3$ film, as expected from Malozemoff's model that we extend for the first time to the case
of ferroelectric antiferromagnets. Polarized neutron reflectometry measurements reveal the presence of a net
magnetic moment within a $\sim$2 nm slab in the BFO at the interface with the CoFeB, reflecting the presence of
uncompensated spins in the BiFeO$_3$, consistent with the observation of exchange bias and enhanced coercivity.
As the ferroelectric domain structure can be easily controlled by an electric field, our results strongly
suggest that the electrical manipulation of magnetization should be feasible at room temperature in BFO-based
exchange-bias heterostructures.

\section*{Acknowledgment}

This study was supported by the E.U. STREP MACOMUFI (033221), the contract FEMMES of the Agence Nationale pour
la Recherche. H.B. also acknowledges financial support from the Conseil G\'en\'eral de l'Essone. The authors
would like to thank H. Jaffr\`es, M. Viret, G. Catalan and J. Scott for fruitful discussions.


\begin{thebibliography}{34}
\bibitem[{\citenamefont{}()}]{DWwidth}
\bibinfo{note}{The domain wall width in BiFeO$_3$ will be discussed in a separate paper}.
\end{thebibliography}
\end{document}